\documentclass{aip-cp}
\usepackage[numbers]{natbib}
\usepackage{rotating}
\usepackage{graphicx}

\begin{document}

\title{Coherent and dissipative transport in a Josephson junction between fermionic superfluids of $^6$Li atoms}

\author[aff1,aff2]{Elettra Neri}
\author[aff1,aff2]{Francesco Scazza}
\author[aff1,aff2]{ Giacomo Roati\corref{cor1}}

\affil[aff1]{Istituto Nazionale di Ottica del Consiglio Nazionale delle Ricerche (INO-CNR), 50019 Sesto Fiorentino, Italy}
\affil[aff2]{LENS and Dipartimento di Fisica e Astronomia, Universit\`{a} di Firenze, 50019 Sesto Fiorentino, Italy}

\corresp[cor1]{Corresponding author: roati@lens.unifi.it}

\maketitle

\begin{abstract}
Quantum systems out of equilibrium offer the possibility of understanding intriguing and challenging problems in modern physics. Studying transport properties is not only valuable to unveil fundamental properties of quantum matter but it is also an excellent tool for developing new quantum devices which inherently employ quantum-mechanical effects. In this contribution, we present our experimental studies on quantum transport using ultracold Fermi gases of $^6$Li atoms. We realize the analogous of a Josephson junction by bisecting fermionic superfluids by a thin optical barrier. We observe coherent dynamics in both the population and in the relative phase between the two reservoirs. For critical parameters, the superfluid dynamics exhibits both coherent and resistive flow due to phase-slippage events manifesting as vortices propagating into the bulk. We uncover also a regime of strong dissipation where the junction operation is irreversibly affected by vortex proliferation. Our studies open new directions for investigating dissipation and superfluid transport in strongly correlated fermionic systems. 
\end{abstract}

\section{INTRODUCTION}

Since more than thirty years the study of quantum effects on transport has been a valuable tool to unveil fundamental properties of quantum matter. While the understanding of transport properties in solids has been a central topic in quantum mechanics, the discovery of new materials is posing new challenges \cite{Datta}. The study of dynamical properties of strongly interacting quantum systems is critical to applications. It is considered the key to the design of new nano-devices with specific functionalities and whose operation is based on quantum-mechanical effects. This technological potential has generated the necessity of developing effective and accurate theoretical tools  to optimize their performance \cite{Ihn}. However, many open questions are still awaiting for a complete solution: the out-of-equilibrium properties of strongly interacting quantum particles are understood only in few scenarios. For these reasons, there is the quest for versatile systems to help design, predict and optimize the performance of these quantum devices. In the last decade, ultracold atomic systems have emerged as ideal quantum platform to investigate condensed matter models and for the investigation of quantum transport 
\cite{Ventra}. In this context, atomic Fermi gases represent an ideal framework, mimicking the complex behavior of electrons in materials \cite{novel}. One main peculiarity that makes such systems so interesting is their unprecedented tunability. In particular, by means of Feshbach resonances it is possible to tune the native attractive two-body interactions from weak to strong. This has allowed to investigate for the first time the smooth crossover from Bose-Einstein condensation of tightly bound molecules to Bardeen-Cooper-Schrieffer (BCS) superfluidity of long-range weakly bound Cooper pairs. Ultracold Fermi gases represent therefore the natural platform to perform transport experiments, extending the cold atoms based quantum simulation into the domain of quantum electronic devices \cite{brantut}. 

In this work, we report on the first observation of the Josephson effect between strongly correlated fermionic superfluids of $^6$Li atoms. The Josephson effect is a paradigmatic example of quantum transport and a pristine quantum phenomenon that reveals the broken symmetry associated with any condensed state \cite{barone,Tinkham}. Josephson junctions are unique physical objects that allow not only to probe quantum phase coherence but also the study of dissipation-driven quantum phase transitions \cite{caldeira}. 

\section{Brief Introduction To The Josephson Effect.}

\begin{figure}[t]
  \centerline{\includegraphics[width=380pt]{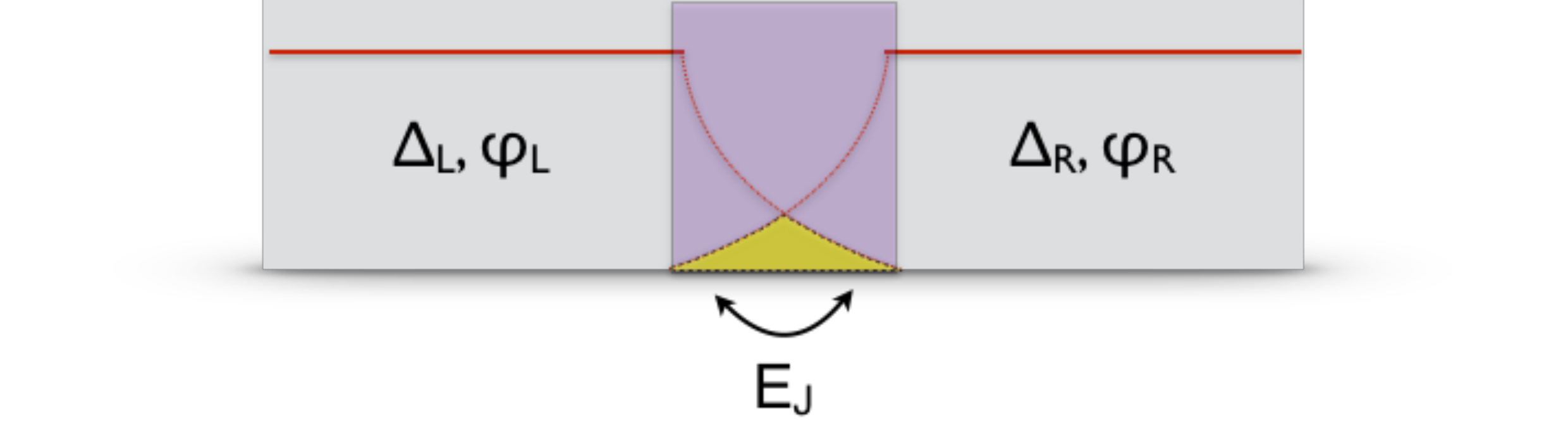}}
  \caption{(a) Sketch of a Josephson junction between two superconducting metals. The two superconductors with order parameters $\Delta_i=|\Delta_i|e^{i\varphi_i}$ (denoted by the red lines) and relative phase $\varphi=\varphi_L-\varphi_R$ are weakly coupled through a thin barrier. Each phase $\varphi_i$ is assumed to be uniform within each superconductor. Despite the exponentially suppression of the order parameter in the insulating region, a finite Josephson coupling $E_J$ may be established by the wavefunctions overlap (yellow region), giving rise to a net supercurrent between the two superconductors.}
  \label{figura0}
\end{figure}

A Josephson junction consists of two superconducting metals in contact through a few nanometer wide insulating barrier, as sketched in Figure~\ref{figura0}. For this configuration, the British physicist B. D. Josephson predicted a remarkable phenomenon \cite{Josephson}: a supercurrent (non dissipative) of Cooper pairs, $I_s=I_C \sin(\varphi)$ should flow between the two superconductors even without an applied voltage, once the relative phase $\varphi=\varphi_L-\varphi_R$ is constant (here, $i=L, R$ indicates the left and the right superconductor, respectively). 
He also proposed that, in the presence of a voltage difference $V$,  $\varphi$ would evolve according to $d\varphi/dt=2eV/\hbar$, leading to an alternating current across the junction $I_s=I_C \sin(2eV/\hbar\times t)$, with amplitude $I_C$ and frequency $\nu=2eV/h$. These two effects are known as dc and ac Josephson effects, respectively. They have been demonstrated in many different experiments in the last decades \cite{barone}. $I_C$ is the maximum critical current that can flow through the junction \cite{Tinkham}. It is related to the Josephson tunneling energy $E_J$, by $E_J\equiv\hbar I_C/2e$, the microscopic quantity that measures how strongly coupled are the phases of the superconductors within the barrier region. Being based on quantum interference, the Josephson effect reveals fundamental aspects of any condensate state. As pointed out by the Nobel Laureate P. W. Anderson: \textit{``The importance of the Josephson effect, then, is that it provides for the first time an instrument which can act like a clamp for a solid or a coercive field for a ferromagnet: it can pin down the order parameter''} \cite{Caianiello}. The observation of the Josephson effect is therefore not limited to superconducting metals but it has been demonstrated in a wide variety of fermionic and bosonic superfluids, from liquid Helium to atomic Bose-Einstein condensates and polaritons \cite{Sukhatme, Hoskinson, Cataliotti, Alb05, Schumm, Lev07, Abbarchi2013}. The essential parameters characterizing a Josephson junction are the differences of condensed pair numbers $\Delta N=N_L-N_R$  and of quantum phases $\varphi$ between the two superconducting reservoirs. These two quantities play the role of conjugate quantum variables as postulated by  P. W. Anderson \cite{Caianiello}, in close analogy with quantum optics where photon numbers and field phases are quasi-conjugate variables. The Josephson relations can be derived from very general considerations in the so-called two-mode approximation introduced by R. Feynman and they are valid for any pair of weakly linked condensates \cite{Feynman}. In particular, they can be obtained as the equations of motion of the \textit{“pendulum Hamiltonian”}: 
\begin{equation}
\label{Josephson-Hamiltonian_0}
H=E_J(1-\cos(\varphi)) + \frac{1}{8}E_C\Delta N^2
\end{equation}
where $E_J$ was defined above, $\Delta N/2 = (N_L-N_R)/2$ is the number of transferred pairs and $E_C=2e^2/C$ is the charging energy with $C$ the electrostatic capacitance of the junction. In the case of the atomic systems investigated in this contribution, $E_J\equiv \hbar I_C$ and  $E_C$ is the capacitive energy due to the interactions between the particles that can be extracted by the equation of state of the trapped gases.

\section{Realizing A Josephson Junction Between Ultracold Fermionic Superfluids.}

\begin{figure}[t]
  \centerline{\includegraphics[width=300pt]{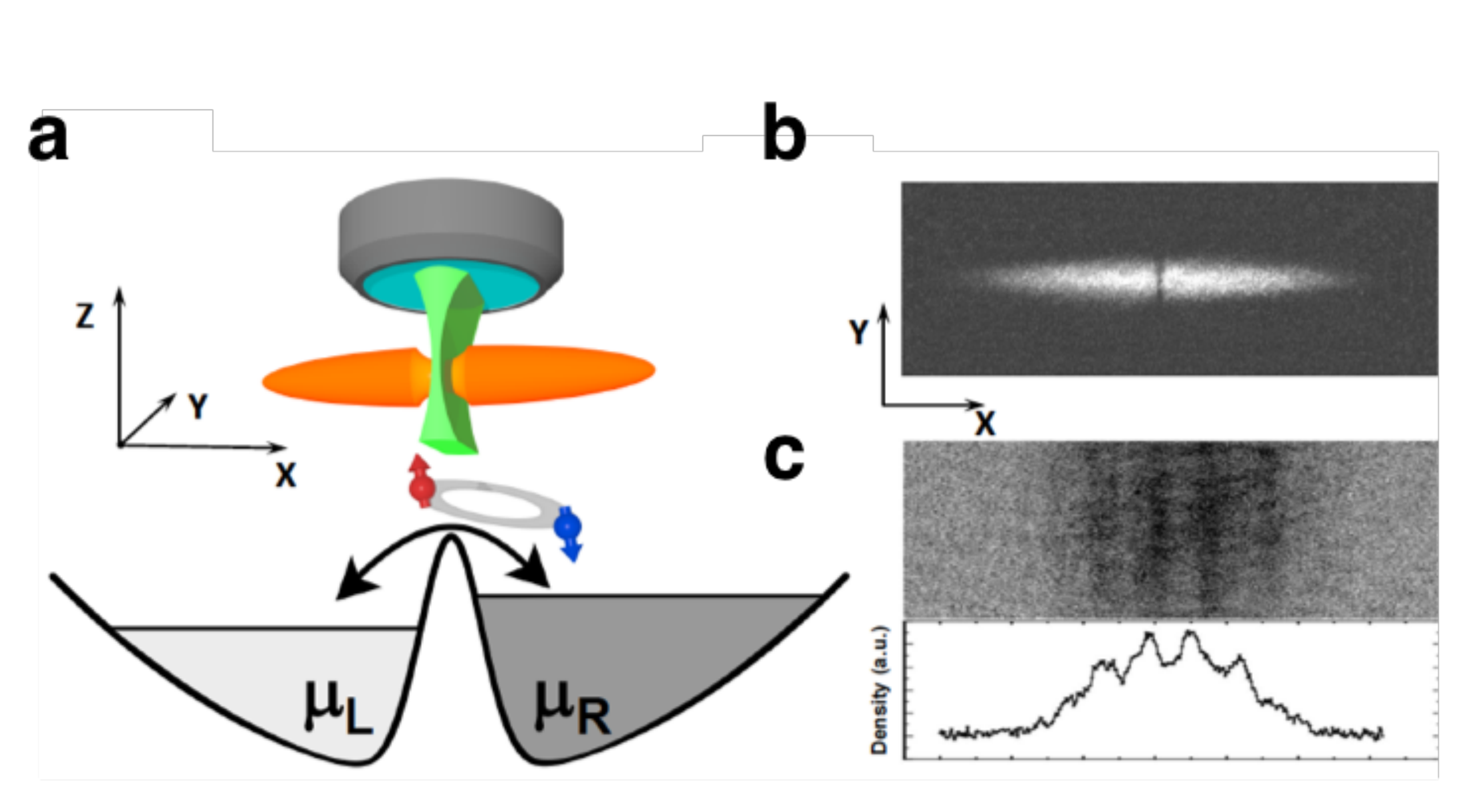}}
  \caption{(a) Sketch of the experimental apparatus. The junction is realized by superimposing on trapped superfluids of $^6Li$ atom pairs an optical barrier. Red and blue arrows indicate the two different spin components forming the fermionic pairs. The dynamics are monitored by recording the number imbalance and relative phase between the two reservoirs via in situ (b) and time-of-flight (c) imaging, respectively (a.u., arbitrary units). From G. Valtolina, et al. \textit{Science} \textbf{350}, 1505 (2015). Reprinted with permission from AAAS.}
  \label{figura1}
\end{figure}

In our experiment, we realize the analogous of a Josephson junction by bisecting fermionic superfluids of $^6$Li atom pairs into two weakly coupled reservoirs by focusing onto the atomic cloud a strongly anisotropic laser beam at 532 nm \cite{valtolina}. This configuration produces a repulsive sheet of light, which splits the cloud in two parts (see Figure~\ref{figura1}(a)). At the trap center, the beam is Gaussian-shaped with a $1/e^2$ waist of $ 2.0 \pm 0.2\,\mu$m and $840 \pm 30\,\mu$m along the $x$ and $y$ directions, respectively. This size is just few times wider than the superfluid coherence length \cite{valtolina}. Since the barrier is almost homogeneous along the $y$ and $z$ directions on the scale of the atomic sample, the total optical potential acting on the pairs can be approximated as: 
\begin{equation}
\label{trapping_potential}
V(\textbf{r})=\frac{1}{2} M(\omega_x^2 x^2+\omega_y^2 y^2+\omega_z^2 z^2) +V_0\, e^{-2x^2/w^2}
\end{equation}
where $w$ is the waist along the x direction, $M=2m$ is the mass of an atomic pair and $V_0$ is the barrier height. This can adjusted in a controlled way by changing the power of the barrier beam.\\
We produce fermionic superfluids of $N \simeq 10^5$ atom pairs by cooling a balanced mixture of the two lowest spin states of $^6$Li, $|F=1/2, m_F=\pm 1/2 \rangle$ (labeled as $|1\rangle$ and $|2\rangle$) to $T/T_F \simeq 0.1$ \cite{Bur14}. Here, $T_F\simeq$ 300~nK is the Fermi temperature, $k_B T_F=E_F=\hbar \,(6N\omega_x \omega_y \omega_z )^{1/3}$, where $k_B$ and $\hbar\!=\!h/(2\pi)$ are the Boltzmann and reduced Planck constants, while $(\omega_x, \omega_y , \omega_z)\simeq 2\pi \times (14, 140, 160)\,$Hz are the trapping frequencies. The optical dipole trap is formed by two laser beams crossing horizontally with an angle of $14^\circ$: the primary beam has a wavelength $\lambda_1=1064$\,nm and a beam waist $w_1\simeq45$\,$\mu$m, while the secondary beam has a wavelength $\lambda_2=1070$\,nm and it is elliptic with beam waists $w_2\simeq45$\,$\mu$m and $w'_2\simeq100$\,$\mu$m. 
The position of the secondary trapping beam can be finely adjusted allowing for the displacement of the total trapping potential center along the axial direction. Interactions between fermions are parametrized by $1/(k_F a)$, where $a$ is the $s$-wave scattering length and $k_F$ is the Fermi wave-vector defined by $E_F = \hbar^2 k_F^2/(2m)$, with $m$ being $^6$Li atomic mass. \\The scattering length between the two spin states is tuned via a broad Feshbach resonance at 832\,G \cite{zurn2013}. The possibility of controlling the interactions close to a Feshbach resonance makes $^6$Li Fermi gases unique systems where to explore many-body phenomena. In particular, it allows for the investigation of the famous BEC-BCS crossover that connects the two paradigmatic regimes of superfluidity: Bose-Einstein condensation (BEC) of composite dimers, and Bardeen-Cooper-Schrieffer (BCS) superfluidity of long-range fermion pairs \cite{novel}. By tuning the s-wave scattering length between fermions from negative to positive, it is possible indeed to vary the size of fermion pairs smoothly, from being much larger than the mean inter-particle spacing (in the BCS limit) up to the small size of a molecular bound state in the BEC limit. The binding energy of the fermion pairs changes in similar fashion: in the BCS limit it is very small (exponentially small with respect to the Fermi energy of the system), while it increases  reaching the binding energy of a molecule in the BEC limit, $E_b\sim\hbar^2/(ma^2)$. At the Feshbach resonance center, the scattering length diverges ($1/k_Fa=0$): the bound state for two isolated particles vanishes while the pair size approaches the mean inter-particle distance, $1/k_F$. More precisely, when $1/k_Fa=1$ the two-body bound state does not play any role and the pairing becomes a pure many-body effect. Here, the chemical potential turns from negative (Bose gas), to positive (Fermi gas). Accordingly, the nature of single-particle excitations changes, and the fermionic nature of the gas becomes relevant (as it is natural for the BCS state). In the crossover regime of diverging scattering length, the gas properties depend on the de Broglie wavelength $\lambda_{dB}\propto 1/\sqrt{T}$ and on the gas density related to the Fermi energy $E_F$. This regime is called unitary or universal. The thermodynamics properties of the gas, such as energy, pressure or compressibility are given by the ones of an ideal, non-interacting Fermi gas multiplied by the Bertsch parameter $\xi$, whose value was found experimentally  $\xi\simeq 0.37$.\\
In the experiments described below, we will explore different regimes of superfluidity across the Feshbach resonance, in particular, (i) a molecular BEC at $1/(k_Fa) \simeq 4$, (ii) a unitary fermionic superfuid at $1/(k_Fa) \simeq 0$, and (iii) a BCS superfluid at $1/(k_Fa) \simeq -0.6$. 

\subsection{Josephson Effect In Fermionic Superfluids Across The BEC-BCS Crossover}

\begin{figure}[t]
\centerline{\includegraphics[width=350pt]{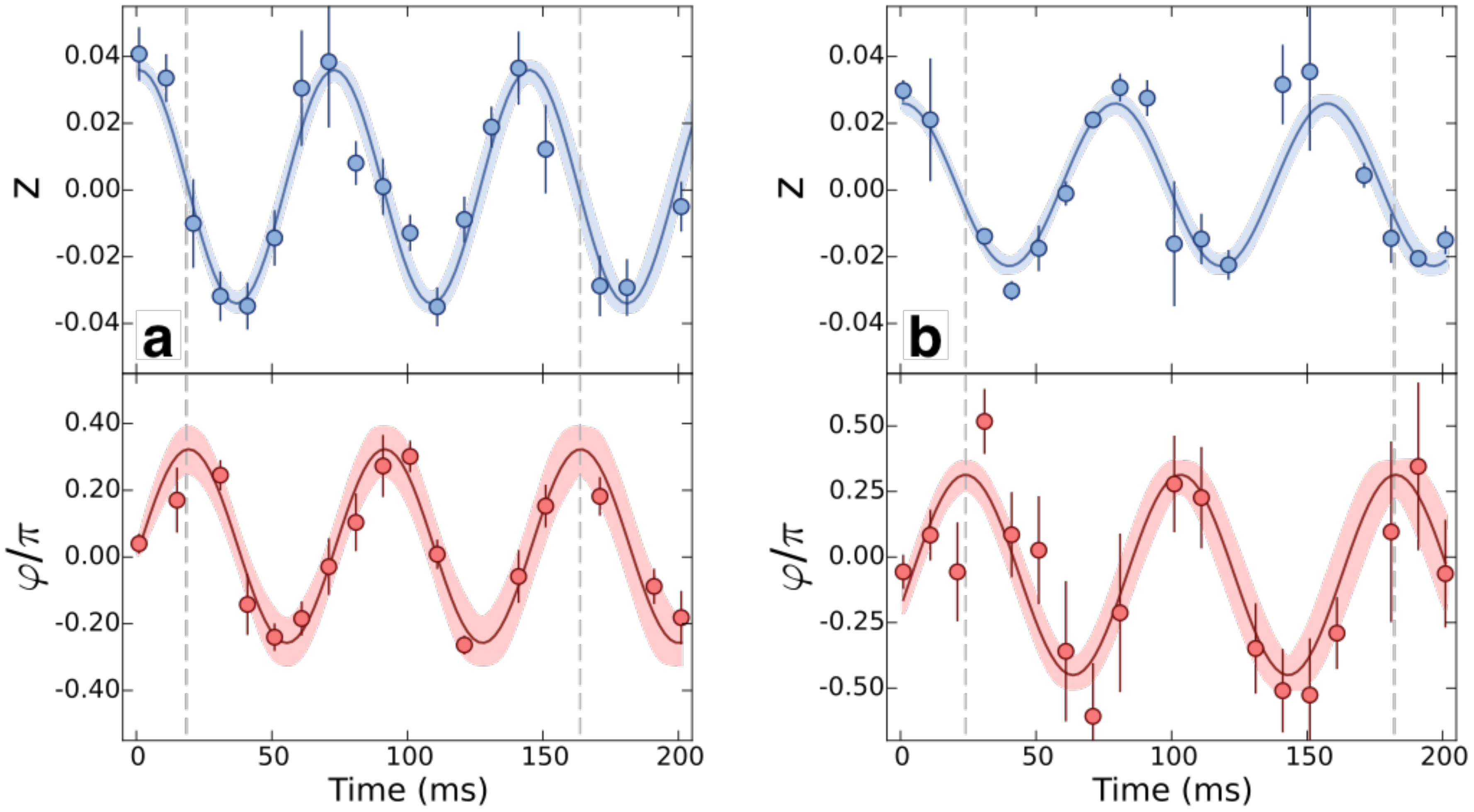}}
\caption{(a) Time evolution of $z(t)$ (blue) and of $\varphi(t)$ (red) in the BEC limit at $1/k_Fa = 4.3$ and $V_0/\mu = 1.0$. Each data point is the average of at least five independent measurements. Error bars are the corresponding standard deviations. Solid lines are fitted to the data with a sinusoidal function. The phase shift $\delta\phi$ between the two oscillations is $1.1 \pm 0.1\, \pi/2$. (b) As in (a), but at unitarity for $V_0/\mu = 1.1$. The relative phase shift is equal to $1.2 \pm 0.2\, \pi/2$. The shaded regions reflect the fit uncertainties. Dashed gray lines indicate the times at which the relative phase reaches its maximum value. From G. Valtolina, et al. \textit{Science} \textbf{350}, 1505 (2015). Reprinted with permission from AAAS.}
  \label{figura2}
\end{figure}
\begin{figure}[t]
  \centerline{\includegraphics[width=380pt]{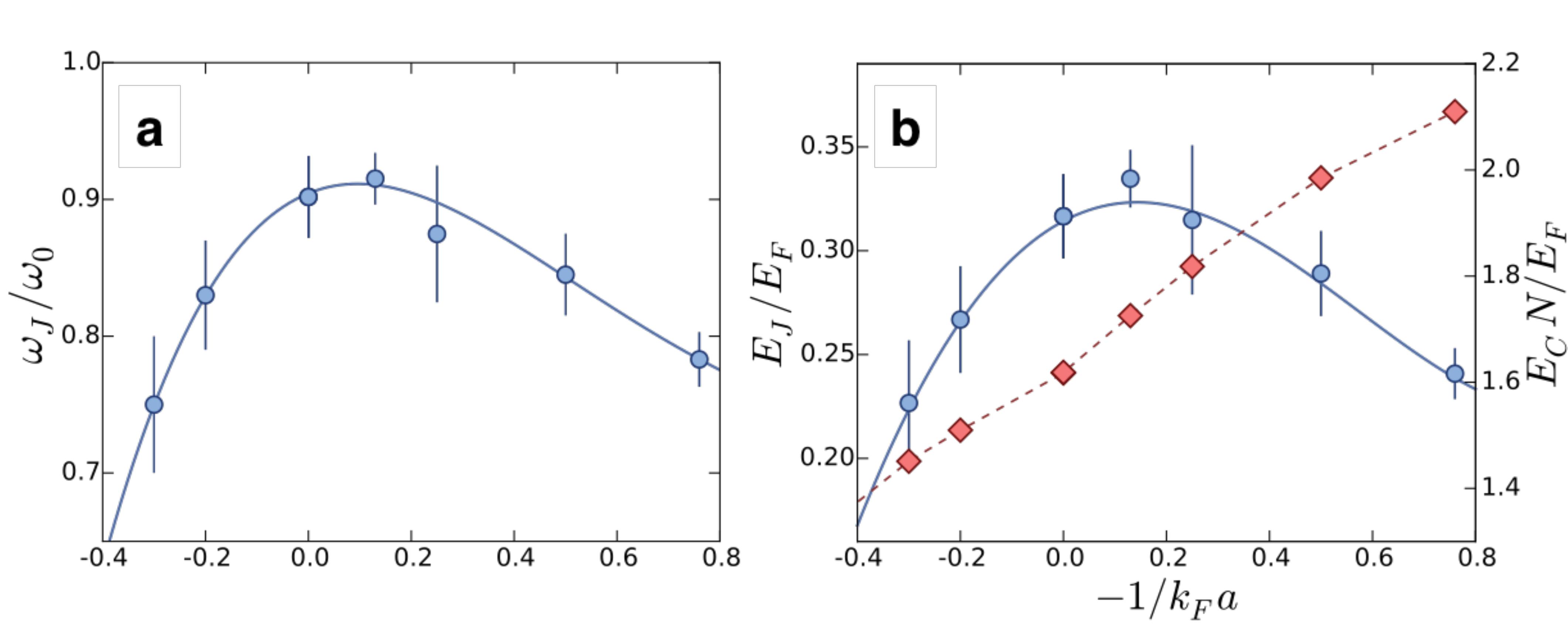}}
  \caption{(a) Josephson plasma frequency $\omega_J$ (blue circles) normalized to the bare trap frequency $\omega_0$ for $V_0/E_F = 1.2 \pm 0.1$, as a function of $1/k_Fa$. The plasma frequency is obtained by fitting $z(t)$ with a sinusoidal function. Error bars indicate the fit uncertainty. (b) Calculated charging energy $E_C$ multiplied for the number of pairs $N$ (red solid diamonds), and Josephson coupling energy $E_J$ (blue circles) normalized to the Fermi energy $E_F$. In both panels, the curves are guides to the eye. From G. Valtolina, et al. \textit{Science} \textbf{350}, 1505 (2015). Reprinted with permission from AAAS.}
  \label{figure3}
\end{figure}

We drive the system dynamics by creating a non-zero initial population imbalance $z_0 = (N_L-N_R)/(N_R+N_L)=\Delta N/N$, where $N=N_L +N_R$ is the sum of the pair populations of the two wells. The imbalance $z_0$ corresponds to a chemical potential difference $\Delta\mu_0 =\mu_L-\mu_R$ across the junction, resembling the applied bias voltage at the terminals of ordinary superconducting junctions. This is achieved by the following procedure. We adiabatically raise the optical barrier just before the end of the evaporation sequence when a superfluid has already formed. We keep the center of the harmonic trap conveniently shifted with respect to the barrier position. Subsequently, by finely adjusting the horizontal position of the focus of the secondary trapping beam, the harmonic trap center is superimposed to the barrier position to obtain an overall symmetric double-well potential. During this procedure $V_0$ is kept well above the value of the gas chemical potential $\mu$ at equilibrium to completely suppress particle tunneling and preserve the desired target imbalance between the two reservoirs. The value of the initial imbalance $z_0$ can be controlled by varying the initial relative displacement of the harmonic trap center. Finally, the dynamics starts by rapidly lowering $V_0$ to the target value in 5\,ms. We monitor the evolution of the population imbalance $z(t)$, equivalent to the current of electron pairs in real Josephson junctions. \\
We initially investigate the regime of small initial imbalances $z_0\sim0.03$, i.e. $\Delta\mu_0/\mu \sim 0.02$ and barrier heights $V_0$ larger than $\mu$ for all the superfluid regimes investigated. The values of the chemical potentials are determined by the measured trap frequencies, atom number and the value of the s-wave scattering length. As we have seen above, the system Hamiltonian may be written in terms of only two macroscopic dynamically conjugate variables, the relative phase and the relative population. In the limit of small excitations and for $E_C, k_BT\ll E_J$, the Hamiltonian~(\ref{Josephson-Hamiltonian_0}) can be further approximated to:
\begin{equation}
\label{JosephsonHamiltonian}
H\simeq E_J\varphi^2/2 + E_C\Delta N^2/8
\end{equation}
The Hamiltonian~(\ref{JosephsonHamiltonian}) rules the dynamics of a non-linear oscillator in which \textit{momentum} $p$ and \textit{position} $x$ are substituted by $\Delta N$ and $\varphi$. The dynamics across the junction is completely determined by the competition of $E_C$ and $E_J$ \cite{Sme97, Zapata, Meier2001}. When tunneling dominates $z$ and $\varphi$ undergo Josephson plasma oscillations, out of phase from each other by $\pi/2$, at a plasma frequency $\omega_J$ that is not depending on the initial imbalance $z_0$ and given by $\hbar \omega_J= \sqrt{E_J E_C}$ where $\hbar$ is the reduced Planck constant. We confirm this expectation by studying the time evolution of both $z$ and $\varphi$ by in-situ and time-of-flight absorption imaging, respectively (see Figure~\ref{figura1}(b)-(c)). The relative phase dynamics is extracted by observing the interference fringes arising from the two expanding clouds after a time of flight of typically 15 ms. In particular, $\varphi$ is extracted by fitting the resulting interferogram by a 2D Gaussian, modulated by a cosine function leaving the phase as a free parameter. While in the BEC side of the resonance, we can measure the relative phase at the same magnetic field at which the dynamics of the imbalance is observed, to determine it in the unitary and BCS regimes, we need to perform a 200-ms fast ramp to the BEC side of the Feshbach resonance. This helps to reduce the detrimental effects of collisions during the expansion, which would completely limit the visibility of the interferogram. In Figure~\ref{figura2}(a)-(b) we present an example of the evolution of $z(t)$ and $\varphi(t)$ for a molecular BEC and a unitary Fermi gas, respectively. In both cases, these quantities oscillate at the same frequency with a measured relative phase shift $\delta\phi = 1.2\pm0.2\,\pi/2$ \cite{valtolina}. This confirms that the imbalance $z$ and the phase $\varphi$ are canonically conjugated variables, probing the macroscopic phase coherence of this strongly interacting system. To understand how the Josephson tunneling depends on the interactions in the BEC-BCS crossover, we measure the Josephson frequency $\omega_J$ for a fixed barrier height $V_0 = 1.2 \pm 0.1\, E_F$, at different values of the interaction strength.  Here we express the barrier peak $V_0$ in units of the Fermi energy $E_F$ of an ideal Fermi gas with the same trap potential and equal atom number. The extracted $\omega_J$ values are shown in Figure~\ref{figure3}(a),
and exhibit a non-monotonic behavior across the BEC-BCS crossover with a maximum around the unitary limit \cite{valtolina}. We can extract $E_J$ from the measured values of $\omega_J$ combined with the computed $E_C$: this is derived from an extended Thomas-Fermi model based on a generalized Gross-Pitaevskii equation which takes into account the proper value of the chemical potential, obtained from QMC (Quantum Monte Carlo) calculations \cite{QMC}. For comparison we plot both $E_J$ and $N\times E_C$ in units of $E_F$  versus the interaction parameter $1/k_Fa$  (see Figure~\ref{figure3}(b)). Whereas $E_C$ increases monotonically moving from the BEC to the BCS regime, $E_J$ reflects the behavior of $\omega_J$, reaching a maximum close to unitarity ($1/k_F a= 0$). In the hydrodynamic limit, $V_0 \ll \mu$, a similar trend has been measured in the determination of the superfluid critical velocity across the whole crossover \cite{Miller2007,spuntarelli}. This quantity depends on the energy spectrum of the system, which in the BEC regime is dominated by sound waves and by pair-breaking mechanisms in the BCS limit. The critical velocity increases while moving towards resonance from the BEC regime and, just before reaching the $1/k_Fa=0$ point, it starts to decrease abruptly, due to the emergence of the single-particle excitation branch featured by  fermionic superfluids. Our measurements are instead carried out in the tunneling regime $V_0 \geq \mu$, and we investigate a coherent effect while avoiding the depletion of the condensate. To understand our observations we consider that, in both the BEC and BCS limits:
\begin{equation}
\label{Josephson-Hamiltonian}
E_J\sim K\times N_0
\label{eq2}
\end{equation}
where $N_0$ is the total number of condensed pairs, and $K$ is a tunneling term that depends on the chemical potential and the barrier properties (geometry, thickness...). In the BEC limit here investigated, the increase of $E_J$ can be understood as due to the increase of the chemical potential $\mu$ while approaching the resonance since almost all the pairs are condensed ($N_0\cong N$).  However, this is not true on the BCS regime where instead $N_0/N \propto\Delta_g/E_F$, where $\Delta_g$ is the superfluid gap, reproducing the $T=0$ Ambeogaokar-Baratoff formula for conventional BCS superconductors \cite{barone}. Toward the BCS limit, $\mu$ does not vary much with $k_Fa$, whereas $N_0$ decreases sharply, causing a net reduction of $E_J$. Our results could provide an alternative way to RF-spectroscopy for the determination of the BCS superfluid gap through tunneling experiments. Interestingly, $E_J$ shows a peak value more shifted towards the BCS limit. On a microscopic level, this should be referred to the increasing robustness of the so-called Andreev-Saint-James bound states towards the unitary limit. These bound states provide the microscopic mechanism for bridging two superconductors separated by a thin barrier \cite{Tinkham}. Accessing the properties of these states in the strongly interacting regime is an active research line from both a theoretical and experimental point of view, also for the condensed matter community \cite{brantut}. 

\subsection{Dissipative Dynamics And Phase Slippage Mechanisms}

\begin{figure}[t]
  \centerline{\includegraphics[width=300pt]{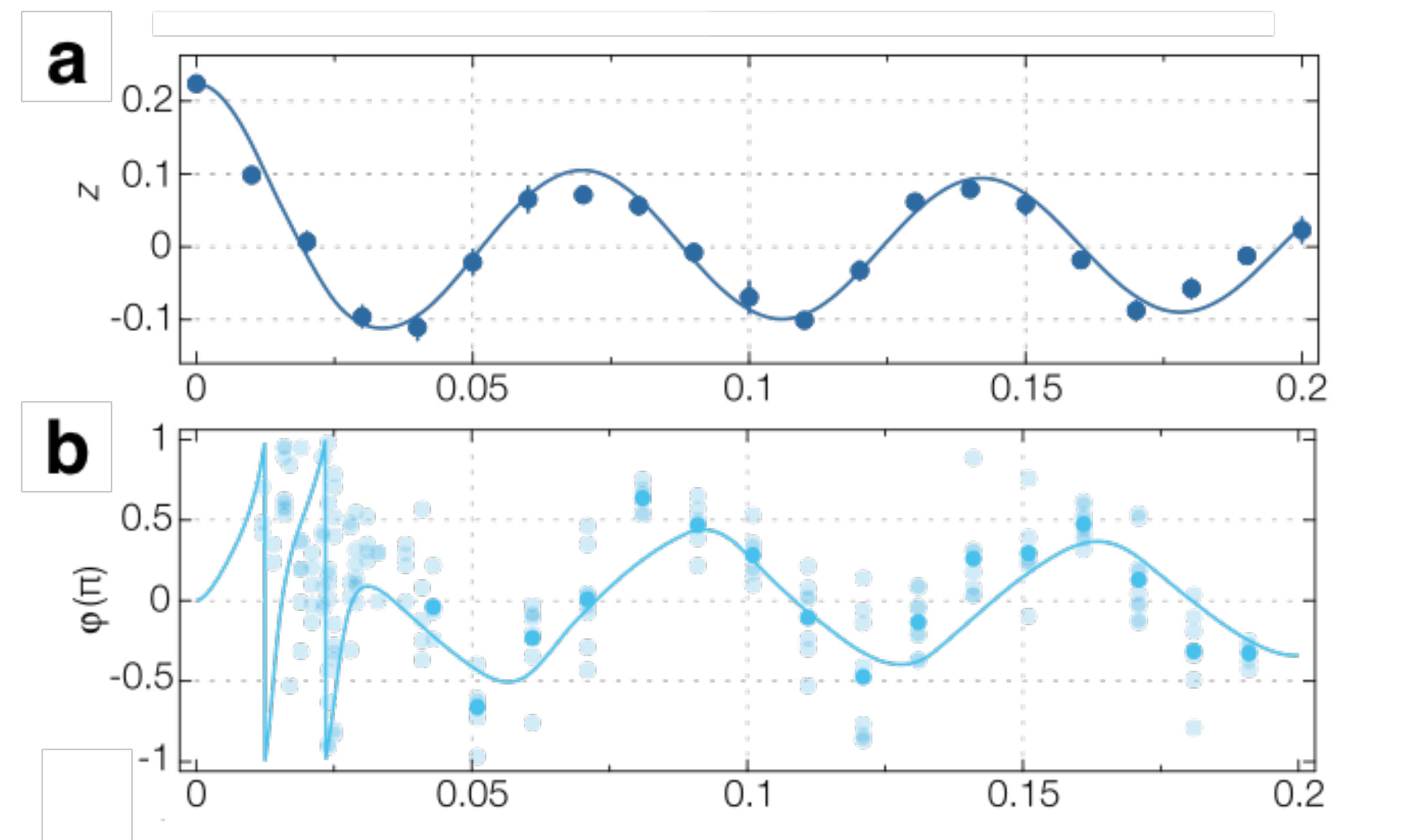}}
  \caption{Population imbalance $z$ (a) and relative phase $\varphi$ (b) evolutions for a molecular BEC at $1/(k_F a)=4.6$ and $V_0/\mu=0.7$. Both solid curves are obtained by a single fit of the measured $z(t)$ with the solution of the coupled circuital model equations (see text). Error bars in panel (a) denote standard errors over at least five independent measurements, while faint circles in panel (b) represent single experimental realizations. Adapted from A. Burchianti et al. \textit{Phys. Rev. Lett.} \textbf{120}, 025302 (2018).}
  \label{figure4}
\end{figure}

\begin{figure}[t]
  \centerline{\includegraphics[width=\columnwidth]{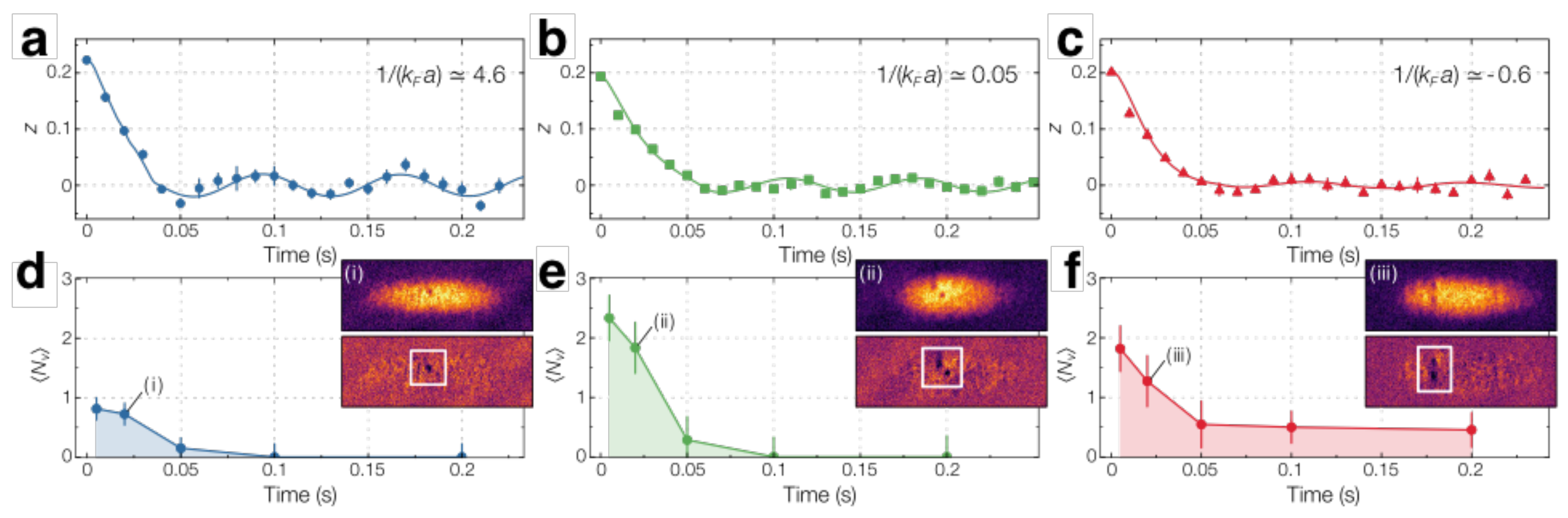}}
  \caption{(a)-(c) Evolution of $z(t)$ with $z_0 \simeq 0.2$ for (a) a molecular BEC at $V_0/\mu \simeq 1$, (b) a unitary Fermi gas at $V_0\mu \simeq 0.9$, and (c) a BCS superfluid at $V_0/\mu \simeq 0.9$. Solid lines are fitted to the data with the solution of the circuital model described in the text. Error bars denote standard errors over at least five measurements. (d)-(f) Evolution of average vortex counts $\left\langle N_v\right\rangle$ for the same experimental conditions as in panels (a)-(c). The error bars are calculated as $\sqrt{\sigma_{Nv}^{2}+1/Z}$, with $\sigma_{Nv}$ the standard deviation of the mean and $Z$ the number of experimental measurements. In the insets (i)-(iii) we show time-of-flight images where vortex defects are clearly visible. For better visualization, residual images are also displayed, where the density distribution of a cloud without excitations has been subtracted. Adapted from A. Burchianti et al. \textit{Phys. Rev. Lett.} \textbf{120}, 025302 (2018).}
  \label{figure5}
\end{figure}

So far, we have investigated the regime of small excitations that characterize the Josephson effect throughout the BEC-BCS crossover. In the following, we will study the regime of larger population imbalance \cite{burk18}. When the initial charging energy $1/8 E_C z_0^2N^2$  exceeds the Josephson coupling $E_J$ the atomic system may enter the macroscopic self-trapping state (MQST) \cite{Alb05, Abbarchi2013, Sme97, Zapata, Abad, Zou2014, Fattori2016}. Here the relative phase increases as $\hbar\varphi(t)\sim E_C z_0 N t$ and it drives small-amplitude oscillations of $z$ around a non-zero value at a frequency $\sim \Delta\mu_0/\hbar$. In our experiment, we instead observe a completely different behavior. In Figure~\ref{figure4}(a)-(b), we show $z(t)$ and $\varphi(t)$ for a molecular BEC with an initial imbalance $z_0 \simeq 0.23$ ($\Delta\mu_0/ \mu \simeq 0.19$) and $V_0/ \mu \simeq 0.7$. We observe that $z$ displays an initial decay alongside a fast variation of $\varphi$ in the range $\left(-\pi,\pi\right)$. 
After this initial dynamics, both $z(t)$ and $\varphi(t)$ oscillate around zero at the same Josephson plasma frequency $\omega<\omega_x$ with a relative phase shift of about $\pi/2$, indicating that Josephson tunneling is indeed re-established after dissipation \cite{burk18}. We point out that the initial variation of $\varphi(t)$ is consistent with the expected running-phase evolution suggesting the emergence of the MQST state. However, in the ideal MQST the imbalance would not evolve from the initial value. The observed irreversible decay of $z$ indicates instead the presence of dissipation mechanisms that lead to the instability of MQST \cite{Lev07, valtolina, Zapata, Meier2001, Zou2014}. The coexistence of running-phase evolution and dissipative flow closely resembles the dynamics of strongly coupled superfluid ${}^4$He reservoirs at $T < T_\lambda$ \cite{Var14, Packard2012}, in which the resistive flow is established by phase-slippage processes and consequent vortex nucleation in the tunnel link. We measure a similar dynamics in the different regimes of superfluidity, as presented in Figure~\ref{figure5}(a)-(c). To understand how resistive and coherent dynamics can coexist, we investigate the microscopic origin of dissipation in our system by monitoring the atomic cloud in time-of-flight after adiabatically removing the barrier potential. We first note that the observed high visibility of the interference pattern at all the interaction strengths rules out the presence of pair-breaking mechanisms as the origin of the dissipative flow. This is in striking difference with respect to ordinary superconducting junctions in which dissipation typically involves the breaking of Cooper pairs \cite{barone, Giaever}. In turn, we observe the initial drop of $z$ to be accompanied by the presence of vortex defects in the superfluid bulk, detected as local density depletions predominantly located within the reservoir at lower initial chemical potential (see Figure~\ref{figure5}(d)-(f)). 

To detect vortex-defects in the crossover superfluids, we slowly sweep the interaction to the BEC side of the resonance to convert all fermionic pairs into tightly bound molecules. This procedure empties out the vortex cores and enhances the defect visibility \cite{MIT_solitons}. We quantify the defects occurrence by performing count-statistics over typically 15 time-of-flight images at various evolution times during the dynamics. In Figure~\ref{figure5}(d)-(f) we show the time evolution of the mean vortex count $\left\langle N_v\right\rangle$ in the same experimental conditions as in Fig.~\ref{figura2}(a)-(c). In all explored superfluid regimes, $\left\langle N_v\right\rangle$ decreases following the behavior of $z$. 
The observed correlated trend of $z(t)$ and $\left\langle N_v\right\rangle\!(t)$ strongly supports the scenario of dissipation driven by vortex-induced phase-slip events \cite{Var14, And66, avenel1985}. 
In this picture the vortex nucleation rate $\gamma$, i.e.~the phase-slip rate, is described by the Josephson-Anderson relation, $\gamma \simeq \dot{\varphi}/(2\pi)  = \Delta\mu/h$ \cite{And66,avenel1985}. For a given $z_0$, $\left\langle N_v\right\rangle$ becomes larger when approaching the weakly attractive regime, in agreement with the increase of $\Delta\mu_0$ from the BEC to the BCS regime. 
Our speculation is confirmed by the fact that once $z(t)$ has dropped below a critical value, the vortex nucleation rate is reduced and pure Josephson dynamics is re-established. Whereas the nucleation rate $\gamma$ is directly linked to $\Delta\mu$, the vortex lifetime is expected to be limited by the presence of the barrier that favors rapid vortex decay into sound waves \cite{Suthar}. Although sound waves must dissipate into heat, we do not observe any related appreciable reduction of the condensed fraction within the measurement timescale. In BEC and unitary superfluids, the number of observed vortices completely decays within $\sim50$\,ms, and no vortices are typically detected in the Josephson regime. 
On the other hand, we observe a sporadic survival of vortices in BCS superfluids even at long evolution times (see Fig.~\ref{figure5}(f)). Even though we lack a complete understanding of this feature, this may result from the combination of the higher nucleation rate and the different effective mass and core-size of vortices in the crossover region \cite{valtolina, Zwi13}. We characterize the nature of these defects by letting them oscillate in the trap in absence of the barrier potential. The trapping potential induces indeed an oscillatory dynamics whose period depends on the features of the crossover superfluids. In the experiment we measure a substantial increase of the vortex period when moving from the BEC to the BCS limit \cite{Zwi13}. This is consistent with the defect being a solitonic vortex, since its period is expected to follow the trend of the chemical potential across the BEC-BCS crossover.\\
Our observations agree with numerical simulations of superfluids connected through a barrier \cite{Abad, Zou2014, piazza}. It is predicted that in a three-dimensional link phase slips arise from vortex rings nucleated within the barrier at the edge of the atomic cloud and oriented perpendicular to the flow \cite{Abad, Var14,piazza}. The combination of low density and inhomogeneity in the junction region causes the vortex core to cross the junction region, shrinking radially in size, thereby leaving behind a phase slip \cite{piazza}. We perform numerical simulations of our system dynamics using the $T=0$ Extended Thomas-Fermi model (ETFM) in both the BEC limit and at unitarity \cite{Forbes_2014}, reproducing qualitatively such vortex dynamics and the observed evolution of $z(t)$. For our large reservoirs and thin barrier, the nucleated vortex rings shrink and escape the barrier region before self-annihilation, explaining our evidence of vortices in the bulk. The vortex shedding leads to the dissipative decay of $z$, and is favored by the multimode character of our junction. Experimentally, we observe defects predominantly oriented along the tighter confining trap axis, i.e.~the imaging line-of-sight (see Fig.~\ref{figure5}(d)-(f)). This is consistent with the instability of vortex rings towards breaking up into vortex lines in a radially asymmetric trap with $\omega_y<\omega_z$ \cite{MIT_solitons}, assisted also by the slow barrier removal prior to imaging. 

\begin{figure}[t]
  \centerline{\includegraphics[width=200pt]{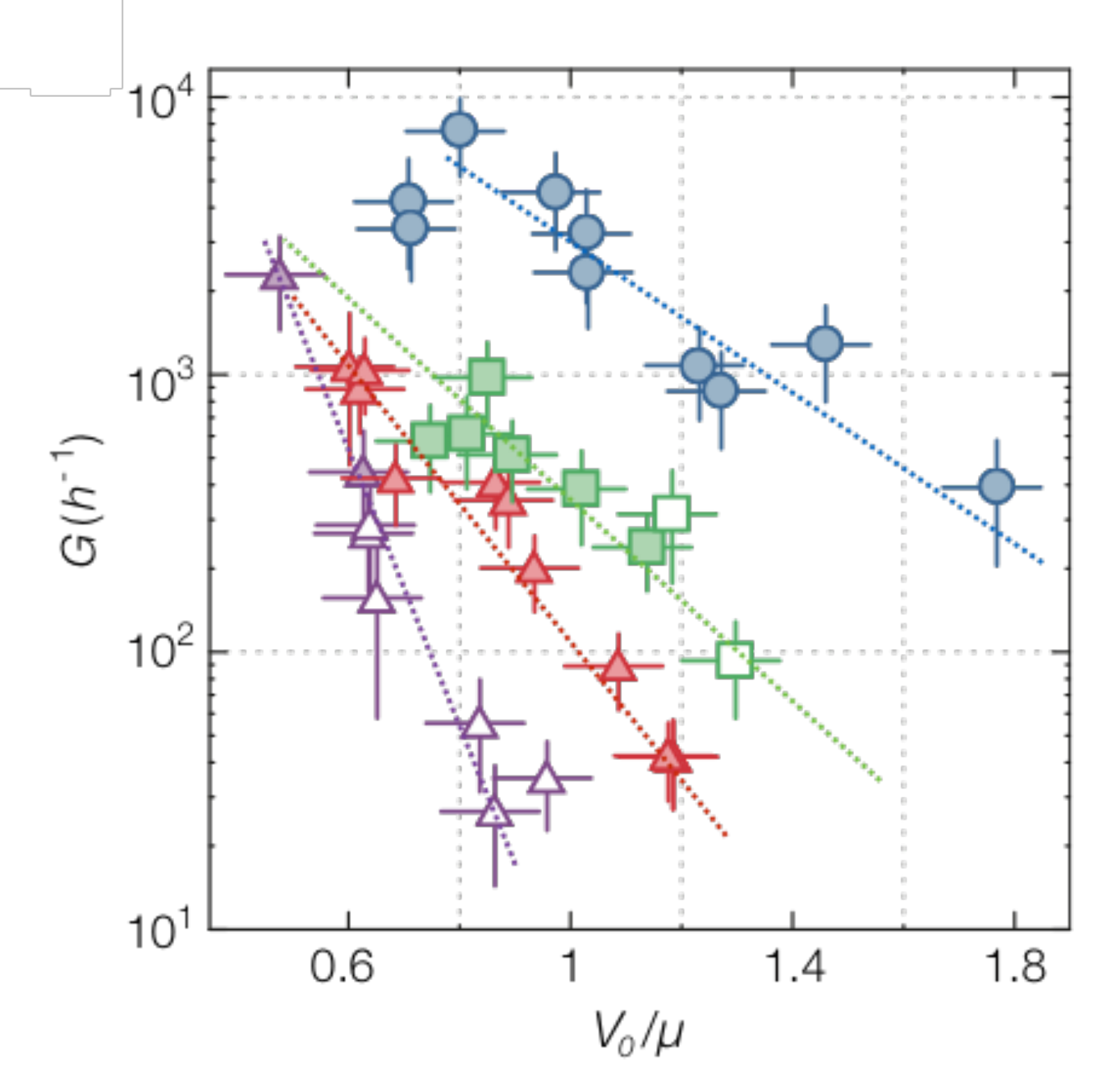}}
  \caption{Conductance $G$ in units of the Planck constant $h$ as a function of the normalized barrier height $V_0/\mu$ for molecular BECs at $1/(k_F a) = 4.6$ (blue circles), unitary Fermi gases (green squares), BCS superfluids at $1/(k_F a) = -0.6$ (red triangles), and an attractive Fermi gas at $1/(k_F a) =-1$ (purple triangles). For filled (empty) symbols, G is obtained through a RSJ-like (RC) circuit model. Dotted lines are guide to the eye. Adapted from A. Burchianti et al. \textit{Phys. Rev. Lett.} \textbf{120}, 025302 (2018).}
  \label{figure6}
\end{figure}

\begin{figure}[t]
  \centerline{\includegraphics[width=350pt]{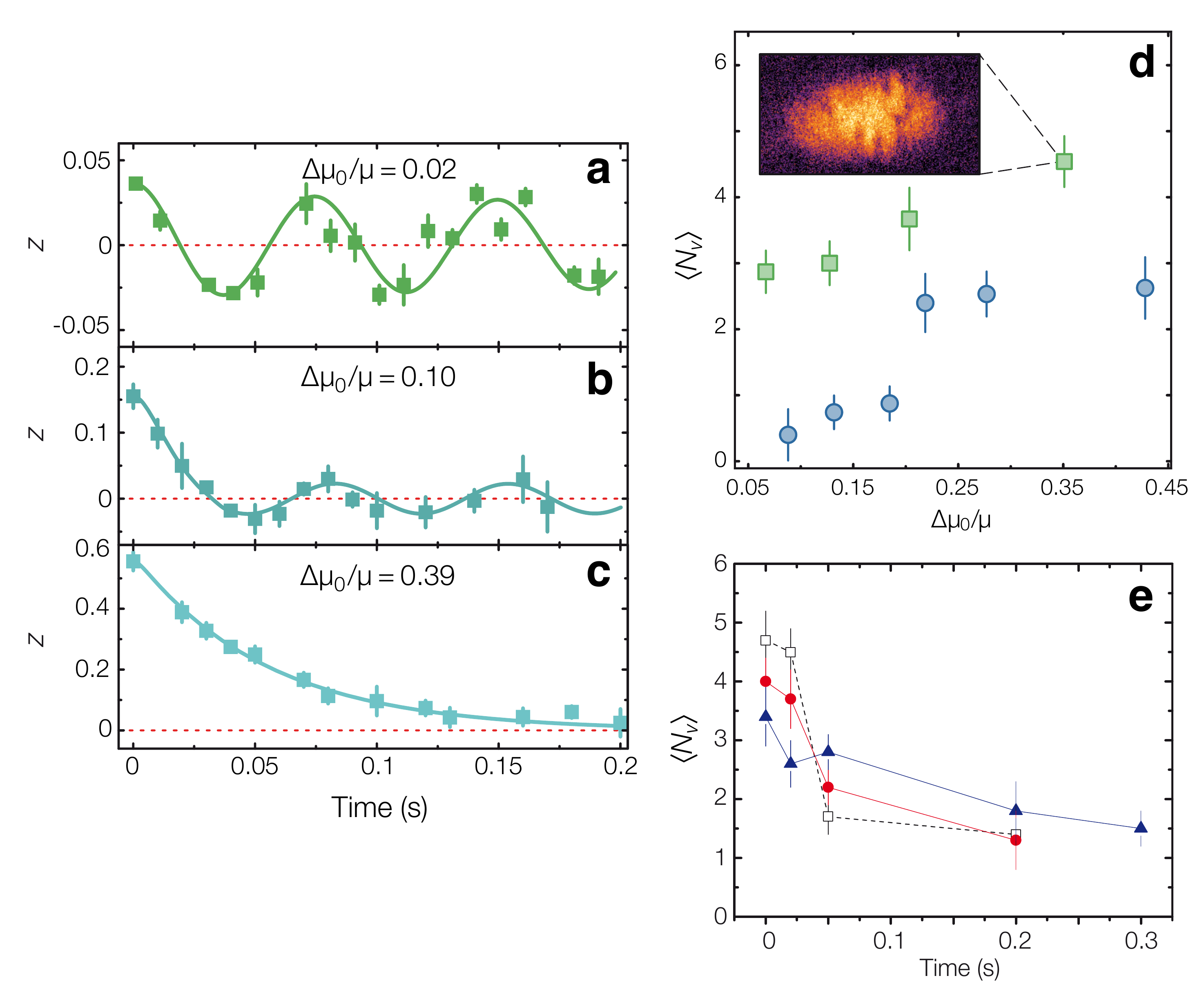}}
  \caption{(a)-(c) Crossover from Josephson to purely dissipative dynamics in a unitary superfluid at $1/k_Fa\simeq 0$ $V_0/\mu\simeq 0.9$. The initial bias potentials $\Delta\mu_0/\mu$ are: (a) 0.02, (b) 0.10 and (c) 0.39. (d) $\left\langle N_v\right\rangle$ as a function of $\Delta\mu_0/\mu$ for the same conditions. Inset: a typical time-of-flight image of an expanding fermionic superfluid for $\Delta\mu_0/\mu\simeq0.4$. Several vortices are clearly visible. (e) Time-evolution of $\left\langle N_v\right\rangle$ for a unitary Fermi gas for $V_0/\mu\simeq 0.9$ for $\Delta\mu_0/\mu \simeq 0.35$ (black opened squares), $\Delta\mu_0/\mu\simeq 0.2$ (red circles) and $\Delta\mu_0/\mu \simeq 0.13$ (blue  triangles). In (d)-(e) error bars are computed as described in the caption of Fig. 5. Dotted lines are guide to the eye. Adapted from A. Burchianti et al. \textit{Phys. Rev. Lett.} \textbf{120}, 025302 (2018).}
  \label{figure7}
\end{figure}
We can describe our junction with an equivalent circuit made of three parallel elements, namely a Josephson weak link with a current-phase relation $I_s=-I_C\sin(\varphi)$, a shunt resistance $R$ and a $LC$ series. This model (\textit{Resistively Shunted Josephson junction, RSJ}) has been originally proposed for describing superconductive Josephson junctions in the presence of dissipation \cite{ barone,Stewart68,McCumber68}. The current/phase time evolutions are given by the following two differential equations \cite{barone}: 
\begin{eqnarray}
L\ddot{k}+ \frac{k}{C} +  R \left(\dot{k} + I_C \sin \varphi \right)= 0
\label{eq1}\\
\hbar \dot{\varphi}=R \left(\dot{k} + I_C \sin \varphi\right),
\label{eq2}
\end{eqnarray}
where we introduce the analogue of the electric charge $k=z N/2$ whose time-derivative gives the particle total current $I$. Eq.~(\ref{eq1}) is the Kirchhoff's voltage law and Eq.~(\ref{eq2}) is the generalized Josephson-Anderson relation for the phase. By numerically solving Eq.~(\ref{eq1}) and Eq.~(\ref{eq2}), we can obtain $z(t)$ and $\varphi(t)$. We fit the measured evolution of $z$ with the calculated one, leaving $R$ and $I_c$ as fitting parameters. In particular, we extract the values of the conductance $G=R^{-1}$ as a function of $V_0/\mu$ for the different superfluids (see Figure~\ref{figure6}). We observe that $G$ decreases while increasing $V_0$, in an almost exponential fashion as expected due to the reduction of the tunneling strength. It is interesting to note that the conductance of an attractive Fermi gas in the normal phase at $1/k_Fa=-1$ lies well below the one of the BCS superfluid, suggesting a different microscopic mechanism dominating the conduction across the barrier, most likely associated with single-particle, incoherent, tunneling. 

Finally, we access a completely different dissipative regime by further increasing the relative imbalance between the two reservoirs. In Figure~\ref{figure7}(a)-(c) we show typical evolutions $z(t)$ for unitary gases ($1/k_Fa\simeq 0$) at three different values of $\Delta \mu_0/ \mu$ with $V_0/\mu\simeq 0.9$. We observe that the increment of $\Delta \mu_0/\mu$ leads a dissipative flow that tends to irreversibly equilibrate the two reservoirs. At large enough imbalance ($\Delta \mu_0/\mu \approx 0.2$) the visibility of the Josephson oscillations is lost. At the same time, we measure  $\left\langle N_v\right\rangle (t)$ varying $\Delta\mu_0/\mu$ (see Figure~\ref{figure7}(d)). The increase of the relative imbalance leads to the increase of $\gamma$ and therefore of $\left\langle N_v\right\rangle$. The disappearance of Josephson oscillations seems to indicate that the coherent coupling between the reservoirs is irreversibly affected by phase-slip proliferation \cite{Var14}. We speculate that the presence of several vortices interacting nearby the barrier creates a local turbulent pad region \cite{Tsatsos, Bulgac}, where the superfluid density is locally suppressed. Nevertheless the saturation of the vortex production rate may arise from vortex reconnections and their mutual interactions \cite{Bulgac,trentoPRX}. This is also corroborated by the observation that the lifetime of vortices depends on the presence of other defects in the condensates: higher $\left\langle N_v\right\rangle$ faster is the decay time (see Figure~\ref{figure7}(e)). On the other hand, the vortices proliferation and their mutual interaction may locally scramble the relative phase $\varphi$ between the two reservoirs, thereby reducing the  coupling between the condensate wave-functions. This corresponds to an effective reduction of the Josephson tunneling energy $E_J$ akin to thermal fluctuations in superconducting junctions \cite{Tinkham}. It is important to note that our observations cannot be ascribed to an increase of the sample temperature, since the condensed fraction in the BEC regime remains above 0.7, limited only by the intrinsic lifetime of the gas. 

\section{CONCLUSIONS}

In conclusion, we have observed the Josephson effect between two fermionic superfluids, weakly coupled through a thin tunneling barrier. For all interaction regimes, we have measured coherent oscillations of both the pair population imbalance and the relative phase, demonstrating that they are canonically conjugate dynamical variables throughout the crossover from the molecular Bose-Einstein condensate (BEC) to the Bardeen-Cooper-Schrieffer (BCS) superfluid regime. Furthermore, we have investigated the emergence of phase-slip-driven dissipation in the dynamics across the junction. Our studies extend the phase-slippage picture of dissipation, typical of liquid helium and certain superconductors, to strongly correlated atomic Fermi gases. Our studies represent the first step towards the investigation of the interplay between dissipative and superfluid transport in strongly correlated fermionic systems.

\section{ACKNOWLEDGMENTS}

We acknowledge all the people that contributed to the results we presented in this work  and in particular: M. Zaccanti, G. Valtolina, A. Amico, A. Burchianti, C. Fort, A. Smerzi, A. Trombettoni, K. Xhani and M. Inguscio. We also acknowledge inspiring discussions with F. Dalfovo, A. Recati, W. Zwerger, T. Giamarchi, F. Piazza and N. Proukakis. We especially acknowledge the LENS Quantum Gases group. This work was supported under European Research Council grant no. 307032 QuFerm2D and it has received funding from the European Union’s Horizon 2020 research and innovation programme under the Marie Sklodowska-Curie grant agreement n. 705269.

%

%
%

\begin{thebibliography}{11}

\bibitem{Datta} S. Datta \textit{Electronic Transport in Mesoscopic Systems} (Cambridge University Press, 1997).

\bibitem{Ihn}T. X. Ihn \textit{Semiconductor Nanostructures: Quantum states and electronic transport} (Oxford University Press, 2010).

\bibitem{Ventra} C.-C. Chien, S. Peotta and M. Di Ventra, \textit{Nature Phys.} \textbf{11}, 998 (2015).

\bibitem{novel} M. Zwierlein \textit{Novel Superfluids Vol.2} (Oxford University Press, 2015).

\bibitem{brantut} S. Krinner, T. Esslinger and J.P. Brantut, J. Phys. Condens. Matter \textbf{29}, 343003 (2017).

\bibitem{barone} A. Barone and G. Patern\`o, \textit{Physics and applications of the Josephson effect} (Wiley, New York, 1982).

\bibitem{Tinkham} M. Tinkham, \textit{Introduction to Superconductivity}, 2nd ed. (McGraw-Hill, New York, 1996).

\bibitem{caldeira} A. O. Caldeira and A. J. Leggett, \textit{Ann. Phys.} \textbf{149}, 374 (1983).

\bibitem{Josephson} B. D. Josephson, \textit{Phys. Letters} \textbf{1}, 251 (1962).

\bibitem{Caianiello} P.  W.  Anderson in \textit{Lectures on the Many-body Problems}, E.R. Caianiello (Eds.)-Elsevier Science (1964).

\bibitem{Sukhatme} K. Sukhatme, Y. Mukharsky, T. Chui, D. Pearson, \textit{Nature} \textbf{411}, 280 (2001).

\bibitem{Hoskinson} E. Hoskinson, Y. Sato, I. Hahn, R. E. Packard, \textit{Nature Phys.} \textbf{2}, 23 (2006).

\bibitem{Cataliotti} F. S. Cataliotti et al.,  \textit{Science} \textbf{293}, 843 (2001).

\bibitem{Alb05} M. Albiez, R. Gati, J. F\"{o}lling, S. Hunsmann, M. Cristiani, and M. K. Oberthaler, \textit{Phys. Rev. Lett.} \textbf{95}, 010402 (2005).

\bibitem{Schumm} T. Schumm et al., \textit{Nature Phys.} \textbf{1}, 57 (2005).

\bibitem{Lev07} S. Levy, E. Lahoud, I. Shomroni, and J. Steinhauer, \textit{Nature} \textbf{449}, 579 (2007).

\bibitem{Abbarchi2013} M. Abbarchi, A. Amo, V. G. Sala, D. D. Solnyshkov, H. Flayac, L. Ferrier, I. Sagnes, E. Galopin, A. Lematre, G. Malpuech, and J. Bloch, \textit{Nature Phys.} \textbf{9}, 275 (2013).

\bibitem{Feynman} R.P. Feynman, R.B. Leighton e M.Sands \textit{The Feynman Lectures on Physics Vol.III}, Addison-Wesley (1965).

\bibitem{valtolina} G. Valtolina, A. Burchianti, A. Amico, E. Neri, K. Xhani, J. A. Seman, A. Trombettoni, A. Smerzi, M. Zaccanti, M. Inguscio, and G. Roati, \textit{Science} \textbf{350}, 1505 (2015).

\bibitem{Bur14} A. Burchianti, G. Valtolina, J. A. Seman, E. Pace, M. De Pas, M. Inguscio, M. Zaccanti, and G. Roati, \textit{Phys. Rev. A} \textbf{90}, 043408 (2014).

\bibitem{zurn2013} G. Z\"urn, T. Lompe, A. N. Wenz, S. Jochim, P. S. Julienne, and J. M. Hutson, \textit{Phys. Rev. Lett.} \textbf{110}, 135301 (2013). 


\bibitem{Sme97} A. Smerzi, S. Fantoni, S. Giovanazzi, and S. R. Shenoy, \textit{Phys. Rev. Lett.} \textbf{79}, 4950 (1997).

\bibitem{Zapata} I. Zapata, F. Sols, A. J. Leggett, \textit{Phys. Rev. A} \textbf{57}, R28(R) (1998).


\bibitem{Meier2001} F. Meier and W. Zwerger, \textit{Phys. Rev. A} \textbf{64}, 033610 (2001).

\bibitem{QMC} G. E. Astrakharchik, J. Boronat, J. Casulleras, S. Giorgini, \textit{Phys. Rev. Lett.} \textbf{95}, 230405 (2005).

\bibitem{Miller2007} D. E. Miller, J. K. Chin, C. A. Stan, Y. Liu, W. Setiawan, C. Sanner, and W. Ketterle, \textit{Phys. Rev. Lett.} \textbf{99}, 070402 (2007).

\bibitem{spuntarelli} A. Spuntarelli, P. Pieri, G. C. Strinati,
\textit{Phys. Rev. Lett.} \textbf{99}, 040401 (2007).

\bibitem{burk18} A. Burchianti, F. Scazza, A. Amico, G. Valtolina, J. A. Seman, C. Fort, M. Zaccanti, M. Inguscio, and G. Roati, \textit{Phys. Rev. Lett.} \textbf{120}, 025302 (2018).

\bibitem{Abad} M. Abad, M. Guilleumas, R. Mayol, F. Piazza, D. M. Jezek, and A. Smerzi, \textit{Eur. Phys. Lett.} \textbf{109}, 40005 (2015).

\bibitem{Zou2014} P. Zou and F. Dalfovo, \textit{J. Low Temp. Phys.} \textbf{177}, 240 (2014).

\bibitem{Fattori2016} G. Spagnolli, G. Semeghini, L. Masi, G. Ferioli, A. Trenkwalder, S. Coop, M. Landini, L. Pezze', G. Modugno, M. Inguscio, A. Smerzi, M. Fattori, \textit{Phys. Rev. Lett.} \textbf{118}, 230403 (2017).

\bibitem{Var14} E. Varoquaux, \textit{Rev. Mod. Phys.} \textbf{87}, 803 (2015).

\bibitem{Packard2012} Y. Sato and R. E. Packard, Rep. Prog. Phys. \textbf{75}, 016401 (2012).

\bibitem{Giaever} I. Giaever, \textit{Phys. Rev. Lett.} \textbf{5}, 464 (1960).

\bibitem{MIT_solitons} M. J. H. Ku, W. Ji, B. Mukherjee, E. Guardado-Sanchez, L. W. Cheuk, T. Yefsah, and M. W. Zwierlein, \textit{Phys. Rev. Lett.} \textbf{113}, 065301 (2014).

\bibitem{And66} P. W. Anderson, \textit{Rev. Mod. Phys.} \textbf{38}, 298 (1966).

\bibitem{avenel1985} O. Avenel and E. Varoquaux, \textit{Phys. Rev. Lett.} \textbf{55}, 2704 (1985).

\bibitem{Suthar} K. Suthar, A. Roy and D. Angom \textit{J. Phys. B: At. Mol. Opt. Phys.} \textbf{47}, 135301 (2014).

\bibitem{Zwi13} T. Yefsah, A. T. Sommer, M. J. H. Ku, L. W. Cheuk, W. Ji, W. S. Bakr, and M. W. Zwierlein, \textit{Nature}  \textbf{499}, 426 (2013).
  
\bibitem{piazza} F. Piazza, L. A. Collins, and A. Smerzi, \textit{New J. Phys.}, \textbf{13}, 043008 (2011).

\bibitem{Forbes_2014} M. M. Forbes and R. Sharma, \textit{Phys. Rev. A} \textbf{90}, 043638 (2014).




\bibitem{Stewart68} W. C. Stewart, \textit{Appl. Phys. Lett.} \textbf{12}, 277 (1968).

\bibitem{McCumber68} D. E. McCumber, \textit{J. Appl. Phys.} \textbf{39}, 3113 (1968).

\bibitem{Tsatsos} M. C. Tsatsos, P. E.S. Tavares, A. Cidrim, A. R. Fritsch, M. A. Caracanhas, F. E. A. dos Santos, C. F. Barenghi, V. S. Bagnato, \textit{Phys. Rep.} \textbf{622}, 1 (2016).

\bibitem{Bulgac} A. Bulgac, M. M. Forbes, and G. Wlazlowski, \textit{J. Phys. B} \textbf{50}, 014001 (2017).

\bibitem{trentoPRX} S. Serafini, L. Galantucci, E. Iseni, T. Bienaim\'e, R. N. Bisset, C. F. Barenghi, F. Dalfovo, G. Lamporesi, and G. Ferrari, \textit{Phys. Rev. X} \textbf{7}, 021031 (2017).

%

\end{thebibliography}

\end{document}